\documentclass[twocolumn,showpacs,preprintnumbers,amsmath,amssymb]{revtex4}
\usepackage{graphicx}% Include figure files
\usepackage{dcolumn}% Align table columns on decimal point
\usepackage{bm}% bold math

\begin{document}

\preprint{submitted to Phys. Rev. B}

\title{
Isotropic photonic band gap and anisotropic structures in transmission spectra of two-dimensional 5-fold and 8-fold symmetric quasiperiodic photonic crystals}

\author{Masashi Hase}
 \email{HASE.Masashi@nims.go.jp}
\author{Hiroshi Miyazaki$^{1}$}
\author{Mitsuru Egashira}
\author{Norio Shinya}
\author{Kenji M. Kojima$^{2}$}
\author{Shin-ichi Uchida$^{2}$}
\affiliation{%
National Institute for Materials Science (NIMS), 1-2-1 Sengen, Tsukuba, 305-0047, Japan
\\$^{1}$Department of Applied Physics, Tohoku University, Aoba, Aramaki-aza, Aoba-ku, Sendai, Miyagi, 980-8579, Japan
\\$^{2}$Department of Superconductivity, The University of Tokyo, 2-11-16 Yayoi, Bunkyo, Tokyo, 113-0032, Japan
}%

\date{\today}% It is always \today, today,

\begin{abstract}

We measured and calculated transmission spectra of two-dimensional quasiperiodic photonic crystals (PCs) based on a 5-fold (Penrose) or 8-fold (octagonal) symmetric quasiperiodic pattern. 
The photonic crystal consisted of dielectric cylindrical rods in air placed normal to the basal plane on vertices of tiles composing the quasiperiodic pattern. 
An isotropic photonic band gap (PBG) appeared in the TM mode, where electric fields were parallel to the rods, even when the real part of a dielectric constant of the rod was as small as 2.4. 
An isotropic PBG-like dip was seen in tiny Penrose and octagonal PCs with only 6 and 9 rods, respectively. 
These results indicate that local multiple light scattering within the tiny PC plays an important role in the PBG formation. 
Besides the isotropic PBG, we found dips depending on the incident angle of the light. 
This is the first report of anisotropic structures clearly observed in transmission spectra of quasiperiodic PCs. 
Based on rod-number and rod-arrangement dependence, it is thought that the shapes and positions of the anisotropic dips are determined by global multiple light scattering covering the whole system. 
In contrast to the isotropic PBG due to local light scattering, we could not find any PBGs due to global light scattering even though we studied transmission spectra of a huge Penrose PC with 466 rods.

\end{abstract}


\maketitle

\section{INTRODUCTION}

Photonic crystals (PCs) have attracted much attention because they are interesting objects of study in physics and because of their potential applications to various optical devices such as micro optical circuits and single-mode light-emitting diodes.\cite{Review1} 
In a PC, a dielectric constant is modulated periodically, and band structures of light are therefore formed. 
This results in the appearance of energy ranges called photonic band gaps (PBGs) in which propagation of the light is forbidden. 
A PBG common to all incident angles of light is called a complete PBG. 

It is also interesting to study optical properties of quasiperiodic photonic crystals based on quasiperiodic lattices. 
One-dimensional quasiperiodic PCs have already been investigated theoretically and experimentally. 
PBGs have been observed in optical multilayers stacked according to a Fibonacci sequence when an incident light was normal to the layers.\cite{Kohmoto87,Gellermann94,Chow93,Hattori94}  
It has also been found that the transmission coefficient $T$ is multifractal and that there is a scaling of the transmission coefficient with increase in Fibonacci sequences at a quarter-wavelength optical thickness when the resonance condition is satisfied. 
Optical reflectivity spectra have been calculated in quasiperiodic multilayers ordered according to the Fibonacci or generalized Fibonacci sequence when the thickness of each layer was much smaller than the wavelength.\cite{Miyazaki89} 
Frequencies at which spiky singularities appear in the spectrum exhibit a self-similar and nested structure.

Recently, two-dimensional quasiperiodic PCs have also been investigated. 
The quasiperiodic PC consists of rods placed perpendicular to the basal plane on vertices of tiles composing a quasiperiodic pattern. 
Complete PBGs have been observed in 5-fold (Penrose pattern),\cite{Bayindir01} 8-fold (octagonal pattern),\cite{Chan98,Cheng99,Jin99} 10-fold,\cite{Jin00} and 12-fold\cite{Jin00,Zoorob00,Zhang01} symmetric quasiperiodic PCs. 
An important feature is that the position and width of the PBG are almost independent of the incident angle of the light. 
This is in contrast to the fact that PBGs are usually anisotropic in periodic PCs. 
Complete PBGs have also been seen in density of states (DOS) of photons even in a small octagonal PC with only 33 rods.\cite{Chan98} 
Thus, it is thought that the existence of complete PBGs depends on the short-range arrangement of rods. 

At present, the following unsolved problems remain. 
The minimum size of a small quasiperiodic PC with a PBG or related structure (a PBG-like dip, as will be shown later) and the degree of anisotropy in the PBG or related structure in a small PC have not been determined in previous studies.
Since the PBGs in the octagonal PC with 33 rods were confirmed by DOS, the degree of anisotropy in the PBGs was not investigated. 
Influence of long-range arrangement of rods on transmission spectra has also not been studied. 
In periodic PCs, it is thought that PBGs are established by long-range lattice periodicity. 
On the other hand, the existence of a PBG formed by long-range arrangement of rods has not been found in a quasiperiodic PC. 
We therefore studied transmission spectra at various incident angles in 5-fold and 8-fold symmetric quasiperiodic PCs with various numbers of rods. 

\section{Methods of experiments and calculation}

The two-dimensional quasiperiodic PC used in our study consisted of dielectric cylindrical rods in air standing normal to the basal plane on vertices of tiles (rhombi or squares) in a 5-fold (Penrose) or 8-fold (octagonal) symmetric quasiperiodic pattern. 
Figure 1 schematically shows rod arrangements of quasiperiodic PCs used in the calculation. 
The circles represent the rods. 
Rods whose $y$ coordinates are the same are defined as belonging to the same layer. 
The numbers of layers and rods in each PC are stated in the legend of Fig. 1. 
The incident direction of the light is parallel to the basal plane. 
An incident angle $\theta$ is defined in Fig. 1. 
In the octagonal PC, besides the 8-fold symmetry, the distribution of rods has mirror symmetry with respect to a line of 22.5 degrees in each 45-degree sector. 
Therefore, the transmission coefficient $T(\theta)$ satisfies the relation $T(\theta) = T(45 - \theta)$. 
In the Penrose PC, besides the 5-fold symmetry, the distribution of rods has mirror symmetry with respect to a line of 36 degrees in each 72-degree sector. 
In addition, any PC with many rods satisfies the relation $T(\theta) = T(180 + \theta)$. 
Thus, the relation $T(\theta) = T(36 - \theta)$ exists. 
It is therefore sufficient to investigate $T(\theta)$ in the range of $0 \le \theta \le 22.5$ and $0 \le \theta \le 18$ degrees for octagonal and Penrose PCs, respectively. 
Two polarization conditions were used in this study: electric fields being parallel to the rods (TM mode) and electric fields being perpendicular to the rods (TE mode). 

The method used for the experiments was as follows.\cite{Hase00} 
To fabricate the PC, a Cu mold was first made by the following procedure. 
Focused laser light was irradiated on the surface of a 1-mm-thick Cu sheet, and holes were drilled at room temperature in air. 
We utilized the 2nd harmonics of Nd: YAG laser light (wavelength, 532 nm). 
Energy per pulse, pulsewidth, and repetition rate were 0.7 mJ, 12 to 14 ns, and 300 Hz, respectively. 
Each hole was drilled by 999 pulses of the laser light. 
The holes were arranged on the vertices of tiles in a quasiperiodic pattern. 
After the Cu mold had been placed in a small vessel, uncured polymer (epoxy resin) was poured into the vessel and the holes were filled with the epoxy resin. 
The epoxy resin was then cured thermally in air. 
Subsequently, the cured epoxy resin together with the Cu mold was detached from the vessel, and only the Cu mold was dissolved in nitric acid, resulting in the formation of an array of epoxy-resin rods, i.e., a PC. 
Each rod of the PC had a radius $r$ of 22 $\mu$m and a length of 500 $\mu$m. 
The real part of the dielectric constant of the epoxy resin $Re(\epsilon)$ was 2.4 at 40 to 200 cm$^{-1}$.

Far-infrared transmission spectra were obtained at room temperature using a Fourier transform spectrometer (Bruker IFS 113v). 
A mercury lamp and a 6- or 23-$\mu$m-thick Mylar film were utilized as a light source and a beamsplitter, respectively. 
A polarizer made of polyethylene with fine grids was placed in front of the PC in sequence from the light source to a detector. 
A slit was placed just before the PC so that the light would irradiate only on the PC. 
Two types of bolometers, both of which were operated at 1.7 K, were used as detectors: one was a Si-composite bolometer covering a spectral region between 20 and 60 cm$^{-1}$, and the other was a Si bolometer covering a spectral region between 40 and 700 cm$^{-1}$. 
The resolution of the measurement was 2 cm$^{-1}$. 
An optical path was evacuated to avoid absorption due to water vapor.

In the numerical calculation, it was assumed that a plane electromagnetic wave was incident upon the system. 
The Maxwell equation was solved by expanding the electromagnetic field in terms of vector cylindrical harmonics.\cite{Stratton41,Yousif88} 
Expansion coefficients were determined from the boundary condition at the surface of each rod by applying the vector addition theorem to the scattered fields from other rods. 
They were used to calculate the distribution of the electromagnetic field and Poynting vector. 
Then the average of the absolute value of the Poynting vector was calculated on the sampling lines shown in Fig. 1. 
The value obtained was normalized by that of the incident plane wave so that it could be regarded as a transmission coefficient. 
Due to the finite sample size, there was additional diffraction of light, which would give the calculated transmission coefficient more than unity in some cases. 
Details of the calculation method will be described in the subsequent paper. 

\section{Isotropic photonic band gap}

An isotropic PBG will be described in this section. 
Figure 2 shows experimental and calculated transmission spectra in the TM mode of the Penrose PCs with $Re(\epsilon) = 2.4$. 
The radius of the rod is $r = 22$ $\mu$m, and a tile side length is $a = 85$ $\mu$m. 
The PC used in the experiment had 31 layers and 724 rods, while that used in the calculation had 33 layers and 186 rods as shown schematically in Fig. 1(a). 
To obtain sufficient intensity of transmitted light, the PC used in the experiment was longer in the $x$ direction than that used in the calculation. 
The experimental and calculated spectra in the figures have been shifted vertically at intervals of about 0.1 and 1, respectively, so they can be clearly seen. 
Similar vertical shifts were performed for other spectra presented in this paper. 
Each experimental spectrum in Fig. 2(a) exhibits a dip around $ka/2 \pi \sim 0.54$, where $k$ is a wave number of light in vacuum. 
This is in agreement with the calculated spectra in Fig. 2(b), where $T$ is nearly 0 around $ka/2 \pi \sim 0.54$ as indicated by the arrow. 
The spatial distribution of energy densities of electromagnetic fields was also calculated. 
The energy density inside the PC around $ka/2 \pi \sim 0.54$ is very small, indicating the existence of a PBG. 
The position and width of the PBG are almost independent of the incident angle, as has been reported for isotropic PBGs in other quasiperiodic PCs.\cite{Bayindir01,Chan98,Cheng99,Jin99,Jin00,Zoorob00,Zhang01} 
Fine structures are observed in the calculated spectra, while they are smeared out in the experimental spectra. 
The absence of fine structures in the experimental spectra is thought to be due to absorption of light inside the rods. 

An isotropic and therefore complete PBG also exists in the octagonal PC. 
Figure 3 shows experimental and calculated transmission spectra in the TM mode of the octagonal PCs with $Re(\epsilon) = 2.4$, $r = 22$ $\mu$m, and $a = 69$ $\mu$m. 
The PC used in the experiment had 17 layers and 873 rods, while that used in the calculation had 19 layers and 173 rods as shown schematically in Fig. 1(b). 
A dip is seen in the experimental spectra at $ka/2 \pi \sim 0.50$ and is in agreement with the calculated spectra, where $T$ is almost 0 in the same $ ka/2 \pi$ as indicated by the arrow. 
As will be shown later, no PBG was observed in the TE mode of the abovementioned Penrose and octagonal PCs.

In quasiperiodic PCs as well as periodic PCs, a large ratio of dielectric constants between rods and a surrounding medium is advantageous for PBG formation. 
In most studies on quasiperiodic PCs, the ratio was larger than 8.9.\cite{Bayindir01,Chan98,Cheng99,Jin99,Jin00,Zhang01} 
Exceptions are the results of 12-fold symmetric quasiperiodic PCs, which have complete PBGs even when the ratio is 2.1.\cite{Zoorob00,Zhang01} 
Our study has shown that complete PBGs exist in Penrose and octagonal PCs with a small ratio of dielectric constants (2.4), suggesting that any quasiperiodic PC can have a PBG in optimum choice of r/a even when the ratio is as small as 2.4.

In order to determine the origin of an isotropic PBG, we calculated transmission spectra of tiny PCs with small numbers of rods. 
Figure 4 shows transmission spectra calculated in the TM mode of Penrose PCs with $Re(\epsilon) = 3.7$, $r = 22$ $\mu$m, and $a = 85$ $\mu$m. 
The PC shown in Fig. 4(a) has 33 layers and 186 rods, while that in Fig. 4(b) has only 6 rods as shown in Fig. 1(c). 
In this investigation, we chose $Re(\epsilon) = 3.7$, because the PBG can be seen more clearly at 3.7 than at 2.4. 
In the calculation of transmission spectra at various incident angles, we usually fixed the sampling line. 
In the case of a 6-rod Penrose PC, on the other hand, the sampling line was rotated around the origin in order for the incident light to be normal to the sampling line. 
This is because a transmission spectrum is strongly affected by the angle between the sampling line and the light direction in such a tiny PC. 
In the 186-rod Penrose PC, a complete isotropic PBG exists at $ka/2 \pi = 0.45 \sim 0.50$. 
In the 6-rod Penrose PC, there is a deep PBG-like dip almost in the same energy range as that of the PBG. 
In addition, the position and width of the PBG-like dip are almost independent of the incident angle. 
Similarly, a tiny octagonal PC with only 9 rods (a central rod and 8 nearest-neighbor rods) exhibits an isotropic PBG-like dip (not shown here). 
According to Conway's theorem, any local pattern of diameter $D$, e.g., a 6-rod Penrose PC, repeats within a distance less than a few $D$'s from the original pattern.\cite{Gardner77} 
As a result, large Penrose and octagonal PCs contain many 6-rod and 9-rod PCs, respectively. 
It is therefore thought that local multiple light scattering within a tiny PC, which causes a PBG-like dip, plays an important role in the PBG formation and that the isotropy of the PBG stems from the isotropy of the PBG-like dip. 
Of course, global multiple light scattering covering a larger region than that of a tiny PC affects PBG formation to some extent, because the PBG-like dip in the tiny PC has finite $T$ and a larger width than that of the PBG, indicating that only the local light scattering within the tiny PC is not sufficient for complete development of a PBG.

In periodic PCs, it is thought that PBGs are established by long-range lattice periodicity. 
Accordingly, the origin of an isotropic PBG in a quasiperiodic PC is different from the origin of a PBG in a periodic PC.
However, Jin et al. have recently shown the existence of PBGs in two-dimensional amorphous photonic materials which do not possess any long-range order but have short-range order.\cite{Jin01} 
A PBG in an amorphous photonic material overlaps the first PBG of a corresponding square-lattice PC. 
Thus, Jin et al. suggested that only short-range order of the square-lattice PC is necessary for formation of the first PBG. 
This is similar to the mechanism of PBG formation in quasiperiodic PCs. 
Further detailed investigation of the relationship between corresponding dips in a local structure and PBGs in a whole system for periodic PCs as well as quasiperiodic PCs is needed. 

Very recently, Miyazaki and Segawa have studied the scattering of plane-wave electromagnetic fields in arrays consisting of dielectric cylindrical rods with various rotational symmetries.\cite{Miyazaki01} 
A PBG was observed in the TM mode and was almost independent of the incident direction of the light. 
This PBG is probably related to the isotropic PBG in a two-dimensional quasiperiodic PC. 

\section{Anisotropic structures in transmission spectra}

Any anisotropic structures in transmission spectra of quasiperiodic PCs have not been reported in previous studies. 
Since anisotropic structures in transmission spectra have been found in our study, they will be described in this section. 
Figure 5 shows experimental and calculated transmission spectra in the TE mode of the same Penrose PCs as those, for which spectra are shown in Fig. 2. 
In contrast to the TM mode, no PBG is observed up to $ka/2 \pi = 1.7$. 
Instead, one or two dips at $ka/2 \pi = 0.50 \sim 0.60$ and one dip at $ka/2 \pi \sim 0.95$ can be seen in each experimental spectrum. 
The positions of these dips depend weakly but clearly on the incident angle. 
We call these dips weakly anisotropic dips. 
In each calculated spectrum, the dips at $ka/2 \pi = 0.50 \sim 0.60$ are reproduced, and there are dips at $ka/2 \pi \sim 1.0$, which probably correspond to the experimental dip at $ka/2 \pi \sim 0.95$. 

Similar results were obtained for the octagonal PCs. 
Figure 6 shows experimental and calculated transmission spectra in the TE mode of the same octagonal PCs as those for which spectra are shown in Fig. 3. 
No PBG is seen up to $ka/2 \pi = 1.4$. 
There is a dip at $ka/2 \pi \sim 0.50$ both in the experimental and calculated spectra, and its position shifts to higher energy with increase in angle.

Figure 7 shows experimental transmission spectra in the TM mode of Penrose and octagonal PCs with $Re(\epsilon) = 2.4$ and $r = 22$ $\mu$m. 
The Penrose and octagonal PCs had $a = 110$ and 100 $\mu$m, 31 and 17 layers, and 562 and 607 rods, respectively. 
There is a dip around $ka/2 \pi \sim 0.60$ in each spectrum of the Penrose and octagonal PCs. 
From calculated spectra (not shown), it is understood that this dip corresponds to an isotropic PBG. 
In addition to the PBG, there are dips seen only at specific angles, i.e., strongly anisotropic dips.
For example, in the Penrose PC, a dip around $ka/2 \pi \sim 0.90$ at 0 degrees and dips around $ka/2 \pi \sim 0.75$ and 1.15 at 18 degrees are not seen at any angle. 
Similarly, in the octagonal PC, a dip around $ka/2 \pi \sim 0.75$ at 0 degrees and dips around $ka/2 \pi \sim 0.70$ and 0.95 at 5.6 to 11.3 degrees are not observed at any angle. 
To the best of our knowledge, this is the first study showing the apparent existence of anisotropy in transmission spectra.

The transmission spectra of PCs with various numbers of rods were calculated. 
Figure 8 shows calculated transmission spectra at 0 degrees of five Penrose PCs with $Re(\epsilon) = 3.7$, $r = 22$ $\mu$m, and $a = 85$ $\mu$m. 
The numbers of rods in the PCs were 6, 16, 31, 186, and 466. 
The PCs with 6, 16, and 31 rods are shown in Fig. 1(c). 
The PCs with 186 and 466 rods have 33 and 70 layers, respectively, and both of these are shown in Fig. 1(a). 
Figure 8(a) shows transmission spectra in the TE mode. 
In the spectrum of the 186-rod PC, there are two dips around $ka/2 \pi = 0.45$ and 0.55. 
These dips correspond to the two dips at $ka/2 \pi = 0.50 \sim 0.60$ in the 186-rod PC with $Re(\epsilon) = 2.4$ in Fig. 5, and show similar weak angle dependence. 
The lower dip around $ka/2 \pi = 0.45$ does not exist in the 6-rod and 16-rod PCs but appears in the 31-rod PC. 
The higher dip around $ka/2 \pi = 0.55$ does not exist in the 6-rod PC but appears in the 16-rod PC. 
The transmission coefficient in both dips decreases with increase in rod number and is very small in the 466-rod PC. 
Figure 8(b) shows transmission spectra in the TM mode.
There is a dip at $ka/2 \pi \sim 0.25$ indicated by the arrow in the 186-rod Penrose PC. 
As is seen in Fig. 4(a), the dip is not observed at an angle greater than 9.7 degrees. 
As shown in Fig. 8(b), the dip does not exist in the 6-rod and 16-rod PCs but appears in the 31-rod PC. 
The transmission coefficient in the dip decreases with increase in rod number and is small in the 466-rod PC. 
As a result, both weakly and strongly anisotropic dips become evident as the rod number increases. 
It should be emphasized that anisotropic dips are not seen in the 6-rod PC, in contrast to the fact that a deep PBG-like dip appears even in the 6-rod PC. 

As can be seen in Fig. 8(a), transmission coefficient is very small in the two dips around $ka/2 \pi = 0.45$ and 0.55 in the 466-rod PC. 
Nevertheless, these dips are not thought to be PBGs. 
A minimum of $T$ in the dips is an order of $10^{-2}$ and is much larger than a minimum of $T$ in the TM-mode PBG in the 466-rod PC, an order of $10^{-6}$. 
The energy densities of the electromagnetic fields were also calculated. 
The ratio of minimum density in the dips to density around a dip edge is about $10^{-2}$ in the TE mode and the ratio of minimum density in the PBG to density around a PBG edge is about $10^{-5}$ in the TM mode. 
Until now, we have not been able to find PBGs other than the isotropic PBGs appearing already in small PCs.

Figure 9 shows calculated transmission spectra at 0 degrees in the TM mode of two Penrose PCs with $Re(\epsilon) = 3.7$, $r = 22$ $\mu$m, and $a = 85$ $\mu$m. 
The lower and upper spectra are spectra of the 186-rod PC (33 layers) in Fig. 1(a) and the 187-rod PC (34 layers) in Fig. 1(d), respectively. 
The lower spectrum and the spectrum of the 186-rod PC in Fig. 8(b) are the same.
These PCs correspond to two different regions inside a colossal Penrose PC with 3946 rods. 
The strongly anisotropic dip at $ka/2 \pi \sim 0.25$ in the lower spectrum does not exist at the same energy in the upper spectrum. 
Therefore, the appearance of anisotropic dips depends on positions of PCs inside the colossal Penrose PC. 
On the other hand, there are isotropic PBGs at $ka/2 \pi = 0.45 \sim 0.50$ in both spectra.
Since the formation of a PBG is mainly due to local light scattering, there is no difference between the PBGs in the two PCs. 

It is thought that the anisotropic dips are determined by global multiple light scattering covering the whole system, because the dips strongly depend on the rod number and rod arrangement. 
This is in contrast to the fact that local light scattering plays an important role in the formation of an isotropic PBG.
We have not been able to find any PBGs formed by global light scattering even though we investigated the transmission spectra of a huge Penrose PC with 466 rods.
This is different from the case of periodic PCs in which the PBG shows its almost complete shape in a system with a finite rod number. 
In periodic PCs, lattice periodicity is a very important factor for the existence of a PBG formed by a Bragg-like multiple scattering mechanism.
If the periodicity is destroyed, then coherence in backscattered waves will be destroyed and so will the PBGs.\cite{Lidorikis00} 
Since quasiperiodic PCs have no lattice periodicity, it is thought that PBGs are not formed by global light scattering. 

We can say that formation of PBGs is easy and difficult in quasiperiodic PCs. 
An isotropic PBG can be formed in a quasiperiodic PC with only a small number of rods. 
As for the dielectric constant, PBGs can be formed in both periodic and quasiperiodic PCs even when the ratio of dielectric constants between rods and the surrounding medium is small. 
However, formation of complete PBGs in periodic PCs is not so easy because of anisotropy. 
On the other hand, a PBG in a quasiperiodic PC is isotropic, and a complete PBG can therefore exist even when the ratio of dielectric constants is small. 
On the other hand, as stated above, formation of PBGs by global light scattering is difficult. 

\section{Summary}

We measured and calculated transmission spectra of two-dimensional quasiperiodic photonic crystals (PCs) based on a 5-fold (Penrose) or 8-fold (octagonal) symmetric quasiperiodic pattern. 
A photonic crystal consisted of dielectric cylindrical rods in air standing normal to the basal plane on vertices of the tiles (rhombi and squares) composing a quasiperiodic pattern. 
A complete isotropic photonic band gap (PBG) appears in the TM mode even when the dielectric constant of the rod is as small as 2.4. 
An isotropic PBG-like dip was observed in tiny Penrose and octagonal PCs with only 6 and 9 rods, respectively. 
It is therefore thought that local multiple light scattering within the tiny PC, which causes the formation of a PBG-like dip, plays an important role in PBG formation. 
In addition to the isotropic PBG, we found dips depending on the incident angle of light. 
This is the first study on anisotropic structures in transmission spectra of quasiperiodic PCs.
Based on the rod-number and rod-arrangement dependence, it is thought that the shapes and positions of the anisotropic dips are determined by global multiple light scattering covering the whole system. 
We have not been able to find any PBGs formed by global light scattering even though we investigated the transmission spectra of a huge Penrose PC with 466 rods. 
This is different from the results for periodic PCs in which the PBG shows its almost complete shape in a system with a finite rod number. 

\begin{acknowledgments}

We are grateful to H. T. Miyazaki, K. Ohtaka, T. Ueta, A. P. Tsai, E. Abe, T. Fujiwara, and A. Yamamoto for their valuable discussions and to H. Takakura and S. Weber for providing data on the coordinates of vertices of the tiles in Penrose and octagonal patterns. 
This work was supported by the budget for Intelligent Material Research from Ministry of Education, Culture, Sports, Science and Technology.

\end{acknowledgments}

\newpage %Just because of unusual number of tables stacked at end

\begin{figure} 
\caption{
Schematic rod arrangements in two-dimensional quasiperiodic PCs. 
Rod positions denoted by circles are the vertices of tiles composing a quasiperiodic pattern. 
5-fold or 8-fold perfect rotational symmetry exists only around the center indicated by the double circle in the Penrose or octagonal pattern. 
The origin of the $xy$ coordinates is defined to be the center. 
The incident angle of light is defined by $\theta$. 
(a) 
A 186-rod Penrose PC (33 layers) is represented by a group of double and open circles, and a 466-rod Penrose PC (70 layers) is represented by all of the circles. 
The side length $a$ of the rhombi is 85 $\mu$m. 
The longer and shorter bars parallel to the $x$ axis are the sampling lines of the Poyinting vector for 186-rod and 466-rod Penrose PCs, respectively. 
Their centers are on the $y$ axis at $y = -446.57$ $\mu$m and -838.13 $\mu$m, and their lengths are 170 and 85 $\mu$m, respectively. 
(b) 
A 173-rod octagonal PC (19 layers) is represented by a group of double and open circles. 
An octagonal pattern is formed by rhombus and square tiles with side length $a = 69$ $\mu$m. 
The bar parallel to the $x$ axis indicated by SL is a sampling line. 
Its center is on the $y$ axis at $y = -377.16$ $\mu$m, and its length is 138 $\mu$m. 
(c) 
A group of double and 5 inside open circles represents a 6-rod Penrose PC, and a group of double, 5 inside open, and 10 closed circles represents a 16-rod Penrose PC. 
The whole set of circles represents a 31-rod Penrose PC. 
The tiles of the Penrose pattern are formed by two kinds of rhombus tiles with side length $a = 85$ $\mu$m. 
The sampling lines SL1 and SL2 are for the 6-rod PC at 0 and 18 degrees, respectively. 
The center of SL1 is at $y = -26.4$ $\mu$m, and SL2 is obtained by rotation of SL1 by 18 degrees around the origin. 
The sampling line SL3 is for the 16- and 33-rod PCs at 0 degrees, and its center is at $y = -106.17$ $\mu$m. 
The length of each sampling line is 68 $\mu$m. 
(d) 
A 187-rod Penrose PC (34 layers) is represented by open circles. 
The side length $a$ of the rhombi is 85 $\mu$m. 
The bar parallel to the $x$ axis indicated by SL is a sampling line whose length is 85 $\mu$m. 
This PC and the 186-rod PC shown in Fig. 1(a) have almost the same sizes and correspond to two different regions inside a colossal Penrose PC with 3946 rods. 
This rod arrangement does not contain a central rod. 
}
\label{F1}
\end{figure}

\begin{figure} 
\caption{
Transmission spectra in the TM mode of Penrose PCs with $Re(\epsilon) = 2.4$, $r = 22$ $\mu$m, and $a = 85$ $\mu$m. 
(a) 
Experimental spectra of a PC with 31 layers and 724 rods. 
The spectra have been shifted vertically at intervals of about 0.1. 
(b) 
Calculated spectra of a PC with 33 layers and 186 rods, which is shown in Fig. 1(a). 
The spectra have been shifted at intervals of 1.
The arrow indicates the position of the PBG. 
}
\label{F2}
\end{figure}

\begin{figure} 
\caption{
Transmission spectra in the TM mode of octagonal PCs with $Re(\epsilon) = 2.4$, $r = 22$ $\mu$m, and $a = 69$ $\mu$m. 
(a) 
Experimental spectra of a PC with 17 layers and 873 rods. 
The spectra have been shifted at intervals of about 0.1. 
(b) 
Calculated spectra of a PC with 19 layers and 173 rods, which is shown in Fig. 1(b). 
The spectra have been shifted at intervals of 1.
The arrow indicates the position of the PBG. 
}
\label{F3}
\end{figure}

\begin{figure} 
\caption{
Calculated transmission spectra in the TM mode of Penrose PCs with $Re(\epsilon) = 3.7$, $r = 22$ $\mu$m, and $a = 85$ $\mu$m. 
The spectra have been shifted at intervals of about 1.
(a) 
A PC with 33 layers and 186 rods shown in Fig. 1(a). 
(b) 
A PC with 6 rods shown in Fig. 1(c). 
}
\label{F4}
\end{figure}

\begin{figure} 
\caption{
Transmission spectra in the TE mode of Penrose PCs with $Re(\epsilon) = 2.4$, $r = 22$ $\mu$m, and $a = 85$ $\mu$m. 
(a) 
Experimental spectra of a PC with 31 layers and 724 rods. 
The spectra have been shifted at intervals of about 0.1. 
(b) 
Calculated spectra of a PC with 33 layers and 186 rods, which is shown in Fig. 1(a). 
The spectra have been shifted at intervals of 1.
}
\label{F5}
\end{figure}

\begin{figure} 
\caption{
Transmission spectra in the TE mode of octagonal PCs with $Re(\epsilon) = 2.4$, $r = 22$ $\mu$m, and $a = 69$ $\mu$m. 
(a) 
Experimental spectra of a PC with 17 layers and 873 rods. 
The spectra have been shifted at intervals of about 0.1. 
(b) 
Calculated spectra of a PC with 19 layers and 173 rods, which is shown in Fig. 1(b). 
The spectra have been shifted at intervals of 1.
}
\label{F6}
\end{figure}

\begin{figure} 
\caption{
Experimental transmission spectra in the TM mode of PCs with $Re(\epsilon) = 2.4$ and $r = 22$ $\mu$m. 
(a) 
A Penrose PC with $a = 110$ $\mu$m, 31 layers, and 562 rods. 
The spectra have been shifted at intervals of about 0.15.
(b) 
An octagonal PC with $a = 100$ $\mu$m, 17 layers, and 607 rods. 
The spectra have been shifted at intervals of about 0.15.
}
\label{F7}
\end{figure}

\begin{figure} 
\caption{
Calculated transmission spectra at 0 degrees of five Penrose PCs with $Re(\epsilon) = 3.7$, $r = 22$ $\mu$m, and $a = 85$ $\mu$m. 
The numbers of rods in the PCs are 6, 16, 31, 186 (33 layers) and 466 (70 layers), which are shown in Fig. 1(a) and (c). 
The spectra have been shifted at intervals of 1.5.
The arrows indicate the positions of angle-dependent dips.
(a) TE mode.
(b) TM mode.
}
\label{F8}
\end{figure}

\begin{figure} 
\caption{
Calculated transmission spectra at 0 degrees in the TM mode of the two Penrose PCs with $Re(\epsilon) = 3.7$, $r = 22$ $\mu$m, and $a = 85$ $\mu$m. 
The lower and upper curves correspond to spectra of the 186-rod PC (33 layers) in Fig. 1(a) and 187-rod PC (34 layers) in Fig. 1(d). 
The upper spectrum has been shifted at an interval of 2.
The arrow indicates the position of an angle-dependent dip. 
}
\label{F9}
\end{figure}

\end{document}